\documentclass[pra,amsmath,amssymb,11pt,tightenlines,a4paper]{revtex4}
\usepackage{amsmath,amssymb}
\usepackage{hyperref}

\begin{document}

\title{Exact flow equation for bound states}
\author{S. Floerchinger}
\affiliation{Institut f\"{u}r Theoretische Physik\\Universit\"at Heidelberg\\Philosophenweg 16, D-69120 Heidelberg}

\begin{abstract}
We develop a formalism to describe the formation of bound states in quantum field theory using an exact renormalization group flow equation. As a concrete example we investigate a nonrelativistic field theory with instantaneous interaction where the flow equations can be solved exactly. However, the formalism is more general and can be applied to relativistic field theories, as well. We also discuss expansion schemes that can be used to find approximate solutions of the flow equations including the essential momentum dependence.
\end{abstract}

\maketitle

\section{Introduction}

The formation of bound states was one of the first problems discussed in quantum mechanics. Although the quantum mechanical formalism for nonrelativistic particles with instantaneous interaction is simple and easy to apply, it is much more difficult to treat the problem in quantum field theory as needed for example for relativistic particles. The Bethe-Salpeter \cite{BetheSalpeter} equation can be used to sum Ladder diagrams but it is difficult to go beyond \cite{Weinberg}. For that reason it is sensible to look for alternative approaches.

In this paper we show how a recently derived exact renormalization group flow equation \cite{FW09} can be used to treat the formation of bound states. As a simple example and to develop the formalism, we consider well known problem of nonrelativistic particles with instantaneous interaction. In this case we can solve the flow equations exactly and we show that our formalism is equivalent to the standard quantum mechanical description although it is based on a field theoretic approach.

In principle, since the flow equation is exact, it can describe also more complicated situations for example with relativistic particles, retarded interactions or at nonzero density. However, one cannot expect to find exact solutions to the flow equations, there. The merit of our formalism is to provide a convenient starting point for approximations. These do not have to rely on the existence of a small coupling constant as in usual perturbation theory and have the potential to capture nonperturbative effects. We discuss a generic scheme for the construction of such approximations. The resulting formalism is quite general and can be seen as a method to resolve the momentum dependence of vertices in renormalization group flow equations.

We mention that flow equations for bound states have been investigated previously. A setup based on Wegner's flow equation for Hamiltonians \cite{Wegner} and the similarity renormalization group \cite{SimilarityRG} has been employed to investigate for example a two-dimensional particle with contact potential \cite{GlazekWilson}. Note that the flow equation of Wegner and the similarity renormalization group break space-time symmetry explicitly. This is in contrast to Wetterich's formulation of renormalization in terms of the effective average action or flowing action $\Gamma_k$ \cite{CWFloweq}. Ellwanger demonstrated that this formulation can be used to investigate bound state formation for relativistic theories \cite{Ellwanger}. By tracing the flow of the four-point vertex, Ellwanger calculates for example binding energies for the Wick-Cutkosky model and finds good agreement with known results in various limits. As a drawback of his approach one might see that properties of the bound states are somewhat hidden in the analytic structure of the four-point vertex. This makes is it hard to refine approximations or to investigate interactions between bound states or spontaneous symmetry breaking. 

These difficulties can be overcome by introducing auxiliary fields for composite operators as will be discussed in more detail below. It is important to adapt the composite fields while solving the flow equation. The reason is that one wants to absorb the essential parts of the four-point vertex by means of an Hubbard-Stratonovich transformation at each step in the flow. A first proposal, how this can be realized with help of a scale-dependent nonlinear field transformation was formulated in ref.\ \cite{GiesWetterich} and used to study bound state formation in the NJL model.  The analysis in the present paper is based on a recently derived, simple, but nevertheless exact flow equation \cite{FW09}. Although the spirit is similar to the formalism proposed in ref.\ \cite{GiesWetterich}, there are some differences in the implementation. 

This paper is organized as follows. In chapter \ref{sec:nonrelativisticmodel} we introduce our notation and the microscopic model for nonrelativistic particles with an instantaneous interaction. In the subsequent section \ref{sec:Partialbosonization} we show how this model can be treated using partial bosonization and an exact flow equation where the Hubbard-Stratonovich transformation is kept fixed, i.e. independent of the renormalization scale $k$. In this approach we can integrate the flow equation exactly. However, this solution relies on some particular features of the nonrelativistic theory and is difficult to generalize to more complicated cases. Therefore we discuss an alternative approach to the problem in section \ref{sec:scaledeppartialbos}. Here we work with a scale-dependent Hubbard-Stratonovich transformation. Again, it is possible to integrate the resulting flow equations exactly in the nonrelativistic case. We argue at the end of the section that this formalism is suited much better for the generalization away from the nonrelativistic and instantaneous approximation. In section \ref{sec:approximationschemes} we suggest approximation schemes and apply our formalism to a nonrelativistic Yukawa potential. Finally, section \ref{sec:generalformalismandapproxschemes} gives a discussion of a somewhat generalized formalism that can be applied for example in relativistic quantum field theory or at nonzero density and we draw some conclusions in section \ref{sec:conclusions}.

\section{Microscopic model for nonrelativistic particles}
\label{sec:nonrelativisticmodel}

Let us consider the problem of two nonrelativistic particles interacting via a potential $V(\vec x-\vec y)$. We will use a quantum field theoretic description using the functional integral. The microscopic action we employ is
\begin{eqnarray}
\nonumber
S_\psi&=&\int_x \psi_1^*(x)\left(i\partial_t+\frac{1}{2M_1}\vec \nabla^2\right) \psi_1(x) \\
\nonumber
&&+ \int_x \psi_2^*(x)\left(i\partial_t+\frac{1}{2M_2}\vec \nabla^2\right) \psi_2(x) \\
\nonumber
&&- \int_{x_1,x_2} \psi_2^*(x_2)\psi_1^*(x_1) \; V(\vec x_1-\vec x_2) \\
&&\times\; \delta((x_1)_0-(x_2)_0)\; \psi_1(x_1)\psi_2(x_2).
\label{eq:microscopicaction2}
\end{eqnarray}
We use here $\int_x=\int_{x_0}\int_{\vec x}=\int dx_0 \int d^3x$. For the Coulomb problem the interaction is of the form
\begin{equation}
V(\vec x_1-\vec x_2) = \frac{-e^2}{4\pi |\vec x_1-\vec x_2|}.
\label{eq:Coulombpot}
\end{equation}
It is useful to change variables of integration according to $x=\eta_1 x_1+\eta_2 x_2$, $y=x_1-x_2$ with $\eta_1=M_1/(M_1+M_2)$, $\eta_2=M_2/(M_1+M_2)$. The interaction term becomes then
\begin{eqnarray}
\nonumber
&&-\int_{x,y} \psi_2^*(x-\eta_1 y)\psi_1^*(x+\eta_2 y) V(\vec y)\\
&& \delta((x_1)_0-(x_2)_0) \psi_1(x+\eta_2 y)\psi_2(x-\eta_1 y).
\end{eqnarray}
It is also useful to introduce the Fourier transformed fields
\begin{eqnarray}
\nonumber
\psi_i(x) &=& \int_p \psi_i(p)\, e^{-ipx}, \quad i=1,2,\quad p x=p_0 x_0-\vec p\vec x,
\end{eqnarray}
with
\begin{equation}
\int_p = \int_{p_0}\int_{\vec p}, \quad \int_{p_0}=\int_{-\infty}^\infty \frac{d p_0}{2\pi}, \quad \int_{\vec p}=\int \frac{d^3p}{(2\pi)^3}.
\end{equation}
We also introduce an abbreviation for a composition of fields $(\psi_1\psi_2)$. Its position and momentum space representation reads
\begin{eqnarray}
\nonumber
(\psi_1\psi_2)(x,\vec y) &=& \psi_1(x_0,\vec x+\eta_2 \vec y) \psi_2(x_0, \vec x-\eta_1 \vec y)\\
&=& \int_{p,q} e^{-ipx} e^{i\vec q\vec y} (\psi_1\psi_2)(p,\vec q),
\end{eqnarray}
with $(\psi_1\psi_2)(p,\vec q)=\int_{q_0}\psi_1(\eta_1 p+q)\psi_2(\eta_2p-q)$. The  conjugate field is
\begin{equation}
(\psi_1\psi_2)^*(p,\vec q) = \int_{q_0} \psi_2^*(\eta_2 p-q) \psi^*(\eta_1 p+q),
\label{eq:}
\end{equation}
where the reversed order of $\psi_1$ and $\psi_2$ is convenient when these are fermionic fields. Obviously, the first variable ($x=(x_0,\vec x)$ or $p=(p_0,\vec p)$) describes the center of mass motion, while the second variable ($\vec y$ or $\vec q$) describes the relative motion of the two particles. The fact that no relative time or energy appears is due to the instantaneous approximation for the interaction. For relativistic systems this point may be different and the generalization of the formalism to that case will be discussed in section \ref{sec:generalformalismandapproxschemes}. 

Using the momentum representation, the interaction term in the microscopic action becomes
\begin{eqnarray}
\nonumber
&&-\int_{x,\vec y} (\psi_1\psi_2)^*(x,\vec y)\, V(\vec y)\, (\psi_1\psi_2)(x,\vec y)\\
&=& -\int_{p,\vec q,\vec q^\prime} (\psi_1\psi_2)^*(p,\vec q)\, V(\vec q-\vec q^\prime)\, (\psi_1\psi_2)(p,\vec q^\prime).
\end{eqnarray}
In the last line we used
\begin{equation}
V(\vec q-\vec q^\prime) = \int_{\vec y} e^{i(\vec q-\vec q^\prime)\vec y} \,V(\vec y).
\end{equation}

In this work we investigate the microscopic action in Eq.\ \eqref{eq:microscopicaction2} using the functional integral formalism. The partition function as a functional of the source field $j$ is given by
\begin{equation}
Z[j] = \int D \psi \; e^{iS_\psi[\psi]-i\int_p\sum_{i=1,2} \{j_i^*(p) \psi_i(p)+c.c.\}}.
\label{eq:partitionfunctnoRk}
\end{equation}
In order to derive an exact flow equation we modify this expression by introducing an infrared cutoff term
\begin{equation}
Z_k[j] = e^{-i W_k[j]} = \int D \psi \, e^{i S_\psi[\psi]+i\Delta S_k[\psi]-i\int \{j^*\psi+c.c.\}},
\end{equation}
where we choose
\begin{eqnarray}
\nonumber
\Delta S_k & = \int_p & \psi_1^*(p)\left[-\frac{1}{2M_1}R_k(\vec p^2)\right]\psi_1(p) \\
&& + \psi_2^*(p)\left[-\frac{1}{2M_2}R_k(\vec p^2)\right]\psi_2(p),
\label{eq:cutoffaction}
\end{eqnarray}
with $R_k\to\infty$ for $k\to0$ and $R_k\to0$ for $k\to0$. The flowing action is now defined as the modified Legendre transform
\begin{equation}
\Gamma_k[\psi] = \int_p \sum_{i=1,2}\{j_i^*(p)\psi_i(p)+c.c.\} -W_k[j]-\Delta S_k[\psi],
\end{equation}
where $\psi(p)=\frac{\delta W_k}{\delta j^*(p)}$ is now the expectation value. The flowing action satisfies the exact flow equation \cite{CWFloweq}
\begin{equation}
\partial_k \Gamma_k[\psi] = \frac{1}{2} \text{STr} (\Gamma^{(2)}[\psi]+{\cal R}_k)^{-1} \partial_k {\cal R}_k.
\label{eq:exactfloweqCW}
\end{equation}
The operation $STr$ is a trace over both continuous and discrete degrees of freedom such as momentum or spin. It also includes a minus sign for fermions. In addition we use in Eq.\ \eqref{eq:exactfloweqCW} the matrix of second functional derivatives
\begin{eqnarray}
&&\Gamma^{(2)}_k(p,q)=\\
\nonumber
&&\begin{pmatrix} \overset{\rightharpoonup}{\delta}_{\psi_1^*(p)} \\ \overset{\rightharpoonup}{\delta}_{\psi_2^*(p)}\\
\overset{\rightharpoonup}{\delta}_{\psi_1(-p)}\\
\overset{\rightharpoonup}{\delta}_{\psi_2(-p)} \end{pmatrix} \Gamma_k[\psi]  \begin{pmatrix} 
\overset{\leftharpoonup}{\delta}_{\psi_1(q)} && \overset{\leftharpoonup}{\delta}_{\psi_2(q)} && \overset{\leftharpoonup}{\delta}_{\psi_1^*(-q)} &&
\overset{\leftharpoonup}{\delta}_{\psi_2^*(-q)} \end{pmatrix}
\end{eqnarray}
and the cutoff matrix
\begin{equation}
{\cal R}_k(p,q) = \begin{pmatrix} \frac{-R_k(\vec p^2)}{2M_1}  && 0 && 0 && 0 \\ 0 && \frac{-R_k(\vec p^2)}{2M_2}  && 0 && 0 \\ 0 && 0 && \frac{R_k(\vec p^2)}{2M_1}  && 0 \\ 0 && 0 && 0 && \frac{R_k(\vec p^2)}{2M_2} \end{pmatrix} \delta(p-q),
\label{eq:secondfunctder}
\end{equation}
with
\begin{equation}
\delta(p) = (2\pi)^4 \delta(p_0) \delta^{(3)}(\vec p).
\label{eq:}
\end{equation}

For large values of $k$ the flowing action approaches the microscopic action
\begin{equation}
\lim_{k\to\infty} \Gamma_k[\psi] = S[\psi],
\end{equation}
while for $k\to0$ the flowing action equals the quantum effective action $\Gamma$ which is the generating functional of one-particle irreducible Feynman diagrams
\begin{equation}
\lim_{k\to0}\Gamma_k[\psi]=\Gamma[\psi].
\end{equation}
We note that the flow equation \eqref{eq:exactfloweqCW} can be written as
\begin{equation}
\partial_k \Gamma_k[\psi] = \tilde \partial_k \frac{1}{2} \text{STr} \ln (\Gamma^{(2)}[\psi]+{\cal R}_k)
\end{equation}
with the formal derivative $\tilde \partial_k$ that applies to the cutoff term ${\cal R}_k$, only.
For reviews of the flow equation method see \cite{ReviewRG,Pawlowski,Metzner,SalmhoferHonerkamp}.

\section{Fixed Hubbard-Stratonovich transformation}
\label{sec:Partialbosonization}

In this section we use a Hubbard-Stratonovich transformation \cite{HS} to calculate the effective bound state propagator. We modify the functional integral over the field $\psi$
 in Eq.\ \eqref{eq:partitionfunctnoRk} by including another integral over the (composite) field $\Phi$
\begin{eqnarray}
\nonumber
Z[j,J] &=& \int D\psi \; e^{i S_\psi[\psi]+i\int_p \sum_i \{j_i^*(p) \psi_i(p)+c.c.\}}\\
& \times & \int D \Phi \; e^{i S_\text{pb}[\psi,\Phi]+i \int_{p,\vec q}\{J^*(p,\vec q)\Phi(p,\vec q)+c.c.\}}
\end{eqnarray}
with
\begin{eqnarray}
\nonumber
S_\text{pb}[\psi,\Phi] &=& \int_{p,\vec q,\vec q^\prime} \left[\Phi^*(p,\vec q) -(\psi_1\psi_2)^*(p,\vec q)\right]\\
&& V(\vec q-\vec q^\prime) \left[\Phi(p,\vec q^{\prime})-(\psi_1\psi_2)(p,\vec q^\prime)\right].
\label{eq:spb1}
\end{eqnarray}
For $J=0$ the functional integral over $\Phi$ contributes only a multiplicative constant to $Z$ which is irrelevant for most purposes. In the combined action $S=S_\psi+S_\text{pb}$ the term quadratic in $\psi$ cancels and we arrive at
\begin{eqnarray}
\nonumber
S &=& \int_p \psi_1^*(p_0-\frac{1}{2M_1}\vec p^2) \psi_1 + \psi_2^*(p_0-\frac{1}{2M_2}\vec p^2)\psi_2\\
\nonumber
&&- \int_{p,\vec q, \vec q^\prime} \left\{\Phi^*(p,\vec q)V(\vec q-\vec q^\prime)(\psi_1\psi_2)(p,\vec q)+c.c.\right\}\\
&&+ \int_{p,\vec q,\vec q^\prime} \Phi^*(p,\vec q) V(\vec q-\vec q^\prime) \Phi(p,\vec q^\prime).
\label{eq:HStransformedaction}
\end{eqnarray}
Note that there is a certain ambiguity in the precise from of the Hubbard-Stratonovich transformation. Linear transformations of the field $\Phi$ are always possible. For the Hubbard-Stratonovich transformed action in Eq.\ \eqref{eq:HStransformedaction} it is now possible to solve some parts of the functional integral.

Let us consider the functional renormalization group equations for this theory when an infrared cutoff of the form in Eq.\ \eqref{eq:cutoffaction} is added for the field $\psi$. For the truncation of the effective action we use a vertex expansion. Due to the nonrelativistic dispersion relation and the instantaneous interaction, the propagator of the field $\psi$ and the Yukawa-like coupling $\sim \Phi^* \psi_1\psi_2+c.c.$ are independent of the scale $k$. Also, no additional interaction vertex involving the field $\psi$ is generated by the renormalization group flow. Indeed, all one-loop Feynman diagrams that contribute to the flow equations of the above quantities consist of a closed tour of particles and vanish in the nonrelativistic few-body limit. This feature is a manifestation of the fact that the nonrelativistic few-body problems decouple in the sence that for example the three-body problem can be solved independent of the four-body problem. More general, the flow equations for correlation functions that determine the $n$-body problem are independent of the correlation functions determining the $n+1$-body problem. For a more detailed discussion of these matters we refer to ref.\ \cite{SFdiss}. As another consequence of the decoupling feature, the flow equation for the propagator of the field $\Phi$ depends only on the propagator of the field $\psi$ and the Yukawa-like vertex but is independent of the higher vertex functions. This implies that the flow equation for the propagator of $\Phi$ can be solved exactly. More concrete, we write the term of $\Gamma_k$ that is quadratic in $\Phi$ as
\begin{equation}
\Gamma_k^{(\Phi,2)} = \int_{p,\vec q,\vec q^\prime} \Phi^*(p,\vec q) P_\Phi(p,\vec q,\vec q^\prime) \Phi(p,\vec q^\prime)
\end{equation}
with a $k$-dependent function $P_\Phi(p,\vec q,\vec q^\prime)$ denoting the inverse propagator. Its flow equation reads
\begin{equation}
\begin{split}
& \partial_k P_\Phi(p,\vec q^\prime,\vec q^{\prime\prime}) =  \tilde \partial_k (-i) \int_{q} V(\vec q^{\prime},\vec q) V(\vec q, \vec q^{\prime\prime}) \\
& \times \frac{1}{(\eta_1 p+q)_0-\frac{1}{2M_1}[(\eta_1 \vec p+\vec q)^2+R_k((\eta_1 \vec p+\vec q)^2)]+i\epsilon}\\
& \times \frac{1}{(\eta_2 p-q)_0-\frac{1}{2M_2}[(\eta_1 \vec p-\vec q)^2+R_k((\eta_1 \vec p-\vec q)^2)]+i\epsilon}.
\end{split}
\end{equation}
We can write this in a symbolic notation as
\begin{equation}
\partial_k P_\Phi(p) = - \tilde \partial_k (V A_k^{-1}(p) V)
\end{equation}
where the ``matrix indices'' $\vec q$ etc.\ and the corresponding integrations are implicit. We use the abbreviation
\begin{equation}
\begin{split}
& A_k^{-1}(p,\vec q,\vec q^\prime) = i \int_{q_0}   \delta^{(3)}(\vec q-\vec q^\prime) \\
& \times \frac{1}{(\eta_1 p+q)_0-\frac{1}{2M_1}[(\eta_1 \vec p+\vec q)^2+R_k((\eta_1 \vec p+\vec q)^2)]+i\epsilon}\\
& \times \frac{1}{(\eta_2 p-q)_0-\frac{1}{2M_2}[(\eta_1 \vec p-\vec q)^2+R_k((\eta_1 \vec p-\vec q)^2)]+i\epsilon}\\
& = \delta^{(3)}(\vec q-\vec q^\prime) {\bigg ( }p_0-\frac{1}{2M_1}[(\eta_1 \vec p+\vec q)^2+R_k((\eta_1\vec p+\vec q)^2)]\\
& \quad \quad -\frac{1}{2M_2}[(\eta_1 \vec p-\vec q)^2+R_k((\eta_1\vec p-\vec q)^2)]+i\epsilon {\bigg )}^{-1}.
\end{split}
\label{eq:akinverse}
\end{equation}
Since the only $k$-dependence in this expression comes from $R_k$, the flow equation for $P_\Phi$ can be integrated directly. For $\Lambda\to\infty$ we obtain
\begin{eqnarray}
P_{\Phi,k}(p)-P_{\Phi,\Lambda}(p) &=&-V A_k^{-1} V.
\end{eqnarray}
For $P_{\Phi,\Lambda}$ we can use
\begin{equation}
P_{\Phi,\Lambda}(p) =  V.
\end{equation}
The propagator
\begin{equation}
G_k(p) = P_{\Phi,k}^{-1}(p) = \left(-V A_k^{-1} V + V\right)^{-1}
\label{eq:propagatorformal}
\end{equation}
describes correlations of two particles with relative momentum $\vec q^\prime$ to two particles with relative momentum $\vec q$. We use here a matrix notation where $\vec q$ and $\vec q^\prime$ are indices and $P_{\Phi,k}(p,\vec q,\vec q^\prime)$ is inverted as a matrix for fixed value of the center of mass momentum $p$ which is conserved.

To understand the physical meaning of $G_k(p)$ we consider field configurations $\Phi$ for which $G_k$ has a pole, or for which 
\begin{equation}
\int_{\vec q^\prime} P_{\Phi,k}(p,\vec q, \vec q^\prime) \Phi(\vec q^\prime) = 0
\end{equation}
holds. In the center of mass frame where $p=(p_0,0,0,0)$ and using the reduced mass $\mu=M_1 M_2/(M_1+M_2)$, this condition can be written as
\begin{equation}
\int_{\vec q^\prime} V(\vec q,\vec q^\prime) \Phi(\vec q^\prime) = \left[p_0-\frac{1}{2\mu}(\vec q^2+R_k(\vec q^2))\right] \Phi(\vec q).
\end{equation}
For $R_k=0$ this is, of course, just the Schr\"odinger equation for the two-body problem. This becomes clear in position space where one obtains
\begin{equation}
\left[p_0-\frac{1}{2\mu}(-\vec \nabla^2+R_k(-\vec \nabla^2))-V(\vec x)\right] \Phi(\vec x)=0.
\end{equation}

In principle, one could obtain the bound state propagator $G_k$ for arbitrary values of the center of mass momentum $p$ by inverting the expression for $P_{\Phi,k}$ according to Eq. \eqref{eq:propagatorformal}. To do that one would first diagonalize the matrix $P_{\Phi,k}$ by finding its eigenvalues $\lambda_i(p)$ and eigenfunctions $\varphi_i(p,\vec q)$. This is a problem of similar complexity as solving the Schr\"odinger equation for the two-particle problem. It is clear that for $p=(E,0,0,0)$ and $k=0$ the spectrum of eigenvalues of $P_\Phi$ contains as many vanishing eigenvalues as bound states exist with energy $E$.

Note, that the on-shell information obtained from solving Schr\"odingers equation (i.e. the binding energies and corresponding eigenfunctions) is not sufficient to determine the bound state propagator $G_k$ uniquely. In particular one can add to $G_k$ a term that is regular as a function of the center of mass momentum $p$ without changing the poles and therefore the on-shell information. It is especially useful to consider the combination
\begin{equation}
\tilde G_k(p) = G_k(p)-G_\Lambda(p) = G_k(p)-V^{-1}.
\end{equation}
This is just the part of $G_k$ that is generated from the flow equation with the ``classical'' part subtracted. Using Eq. \eqref{eq:propagatorformal} this can be written as
\begin{equation}
\tilde G_k(p) = (A(p)-V)^{-1}.
\end{equation}
In this representation it is particularly clear that the poles of $\tilde G_k$ correspond to the solutions of Schr\"odingers equation for two particles. 

Moreover, it is clear how to diagonalize this propagator, at least for $R_k=0$. To that end we note that the inverse propagator has the position space representation
\begin{equation}
\tilde G_{k=0}^{-1}(p) = p_0-\frac{1}{2(M_1+M_2)}\vec p^2-\frac{1}{2\mu}(-\vec \nabla_y^2)-V(\vec y)
\end{equation}
which is diagonalized, of course, by the solutions of the stationary Schr\"odinger equation
\begin{equation}
\tilde G_{k=0}^{-1}(p) g_n(\vec y)=(p_0-\frac{1}{2(M_1+M_2)}\vec p^2-E_n) g_n(\vec y).
\end{equation}
Here, $n$ is a combined index that labels all quantum numbers. For a spherical symmetric potential this includes the radial quantum number as well as those for angular momentum. In the basis 
\begin{equation}
\Phi(p,\vec y) = \sum_n \phi_n(p) g_n(\vec y)
\end{equation}
we can write
\begin{equation}
(\tilde G_{k=0}(p))_{n m} = (p_0-\frac{1}{2(M_1+M_2)}\vec p^2-E_n) \delta_{n m}.
\end{equation}
For completeness we note the explicit form of the Yukawa coupling term in the effective action in this basis
\begin{equation}
\Gamma_k^{\Phi\psi\psi} = \sum_n \int_{p,\vec y} \left\{\phi^*_n(p) g^*_n(\vec y) V(\vec y) (\psi_1\psi_2)(p,\vec y)+c.c.\right\}.
\end{equation}
For a constant cutoff function $R_k=k^2$ the basis of functions $g_n$ diagonalizes also $\tilde G_k$ for nonzero $k$. However, the situation becomes more complicated for other choices of $R_k$ where the basis would depend on the center of mass momentum $p$ and the flow parameter $k$. 

At this point we could in principle undo the Hubbard-Stratonovich transformation and thus go back to our original formulation of the theory in terms of the fields $\psi$. This can be done by solving the field equation 
\begin{equation}
\frac{\delta \Gamma_k}{\delta \Phi}=0
\end{equation}
for the field $\Phi$ as a functional of $\psi$. Plugging this solution $\Phi[\psi]$ into the action leads to
\begin{equation}
\Gamma_k^{(\psi)} = \Gamma_k[\psi,\Phi[\psi]]. 
\end{equation}
In general, the dependence of $\Gamma_k$ on $\Phi$ is complicated and undoing the Hubbard-Stratonovich transformation leads to complicated interaction vertices for the field $\psi$. However, it is quite easy to calculate the effective vertex with two incoming and two outgoing $\psi$-particles. We note that the only contribution to this interaction is given by a tree level diagram involving the propagator $G_k$. The corresponding term in $\Gamma_{k=0}^{(\psi)}$ is
\begin{equation}
\begin{split}
& \Gamma_{k=0}^{(\psi,4)} = -\int_{p,\vec y_1,\vec y_2} (\psi_1\psi_2)(p,\vec y_1)^* g_n(\vec y_1) V(\vec y_1) \\
& \times \frac{1}{p_0-\frac{1}{2(M_1+M_2)}\vec p^2-E_n} V(\vec y_2) g^*_n(\vec y_2) (\psi_1\psi_2)(p,\vec y_2)\\
& - \int_{p,\vec y} (\psi_1\psi_2)^*(p,\vec y) V(\vec y) (\psi_1\psi_2)(p,\vec y).
\end{split}
\label{eq:efffourfermioninteraction}
\end{equation}
This expression has a simple interpretation. The first term is the contribution from bound state exchange processes, while the second term is just the classical term. For an electromagnetic Coulomb interaction, the second term can be seen as a contribution from photon exchange processes.

Let us summarize what we have done in this section. Starting from the microscopic model in Eq.\ \eqref{eq:microscopicaction2} we performed a Hubbard-Stratonovich transformation and introduced the bilocal field $\Phi$. In this ``partially bosonized'' language it was possible to find a solution of the flow equation for the propagator of the field $\Phi$ in a closed form. Note however, that this solution relies on some features particular to the nonrelativistic few-body problem. First, the description of the interaction between particles in terms of a instantaneous interaction potential $V(\vec x_1-\vec x_2)$ is usually not possible for relativistic problems. In relativistic quantum field theory, interactions are mediated by exchange particles such as the photon. In momentum space the resulting interaction term has nontrivial frequency- and momentum dependence and is subject to renormalization group modifications for example due to a ``running'' fine-structure constant $\alpha$ or charge $e$. These features make it difficult to absorb the interaction term by an appropriate Hubbard-Stratonovich transformation. Even if this was possible, solving the flow equations would be more difficult than in the nonrelativistic case presented above. Both the propagator for the field $\psi$ and the Yukawa interaction $\sim \Phi^* \psi_1\psi_2+c.c.$ might have non-vanishing flow equations. These equations as well as the flow equations for the propagator of $\Phi$ would depend on higher-order vertex functions such that no solution in a closed form can be expected.

Although one cannot expect to find analytic solutions by transferring the above calculation to a relativistic field theory, one might ask whether it can be helpful for finding approximate solutions. One could make a truncation of the flowing action $\Gamma_k$ for example in terms of a derivative expansion and consider the flow equations in the theory-subspace spanned by the finite number of operators included in this truncation. Indeed, we will argue below that such truncations can lead to good approximate solutions. The formulation in the present section has one drawback for approximate solutions, however. As already discussed above, interactions in relatvistic field theories are usually mediated by exchange fields such as the photon. The renormalization of the corresponding couplings and propagators can be calculated most efficient in a formulation of the theory which directly takes the exchange fields as propagating fields into account. Other formulations where these fields (as for example the photon) are ``integrated out'' might be equivalent in principle, but are usually much harder to treat by approximate methods. The reason is that the essential momentum- and frequency dependence is often hidden in such formulations.

Transfered to the example of a nonrelativistic theory it would be useful to have a formulation where the two terms in Eq.\ \eqref{eq:efffourfermioninteraction} are treated in different ways. While the term in the first line (the contribution from bound state exchange processes) is treated most efficient in terms of the Hubbard-Stratonovich field $\Phi$, this is different for the second term. For a relativistic field theory this corresponds to the contribution from photon or other exchange processes and it is therefore most efficient to write it in terms of this exchange field. In other words, for the nonrelativistic theory, the Hubbard-Stratonovich transformation should be constructed such that the effective action $\Gamma_{k=0}$ contains as an explicit contribution only the classical term (the second term in Eq. \eqref{eq:efffourfermioninteraction}). All additional terms should be described by bound state exchange processes. For the flowing action one can use the same prescription. To that end it is necessary to work with a scale-dependent Hubbard-Stratonovich transformation, however. An exact flow equation that can be used for this purpose was derived in \cite{FW09} and will be discussed in the next section.

\section{Scale dependent Hubbard-Stratonovich transformation}
\label{sec:scaledeppartialbos}
In this section we investigate the bound state problem using a $k$-dependent version of the Hubbard Stratonovich transformation. Instead of $S_\text{pb}$ in Eq.\ \eqref{eq:spb1} we employ
\begin{equation}
\begin{split}
& S_\text{pb}=\int_p\left[\Phi^*(p)-(\psi_1\psi_2)^*(p) V Q_\Lambda^{-1}(p)\right]\; Q(p)\\
& \left[\Phi(p)-Q_\Lambda^{-1}(p) V (\psi_1\psi_2)(p)\right].
\end{split}
\label{eq:Spb2}
\end{equation}
Again we use a matrix notation where the summation over the index $\vec q$ etc. is left implicit. In Eq.\ \eqref{eq:Spb2} the matrix $Q(p)$ is $k$-dependent, while $Q_\Lambda(p)$ equals $Q(p)$ for $k=\Lambda$ but is independent of $k$. We choose
\begin{equation}
\begin{split}
& \lim_{\Lambda\to\infty} Q_\Lambda(p,\vec q,\vec q^\prime) = \infty \;\delta^{(3)}(\vec q-\vec q^\prime),\\ 
& \lim_{\Lambda\to\infty} Q_\Lambda^{-1}(p,\vec q,\vec q^\prime)=0.
\end{split}
\end{equation}
In this limit the combined action $S=S_\psi+S_\text{pb}$ becomes
\begin{eqnarray}
\nonumber
S &=& \int_p \, \psi_1^*\left(p_0-\frac{1}{2M_1}\vec p^2\right)\psi_1 + \psi_2^*\left(p_0-\frac{1}{2M_2}\vec p^2\right)\psi_2\\
\nonumber
&& - \int_{p,\vec q,\vec q^\prime} (\psi_1\psi_2)^*(p,\vec q) V(\vec q-\vec q^\prime) (\psi_1\psi_2)(p,\vec q^\prime)\\
\nonumber
&& - \int_{p,\vec q,\vec q^\prime} \left\{ \Phi^*(p,\vec q) V(\vec q-\vec q^\prime) (\psi_1\psi_2)(p,\vec q^\prime)+c.c.\right\}\\
&& + \int_{p,\vec q, \vec q^\prime} \Phi^*(p,\vec q) Q(p,\vec q,\vec q^\prime) \Phi(p,\vec q^\prime).
\label{eq:combinedaction2}
\end{eqnarray}
The field $\Phi$ is very ``massive'' for large $\Lambda$. The action in Eq.\ \eqref{eq:combinedaction2} is the starting point for the flow of the functional $\Gamma_k[\psi,\Phi]$ for $k=\Lambda$. In ref.\ \cite{FW09} we showed that the flowing action for a scale dependent Hubbard-Stratonovich transformation of the form in Eq. \eqref{eq:Spb2} satisfies the exact flow equation
\begin{eqnarray}
\nonumber
\partial_k \Gamma_k &=& \frac{1}{2} \text{STr} (\Gamma_k^{(2)}+{\cal R}_k)^{-1} (\partial_k {\cal R}_k - {\cal R}_k (\partial_k Q^{-1}){\cal R}_k)\\
&&-\frac{1}{2} \Gamma_k^{(1)} (\partial_k Q^{-1}) \Gamma_k^{(1)}+\gamma_k
\label{eq:exactflow}
\end{eqnarray}
where $\gamma_k$ is a field independent constant that is irrelevant for most purposes and will be dropped from here on. The flow equation \eqref{eq:exactflow} holds for fixed ($k$-independent) field $\Phi$. For our purpose it is useful to perform a $k$-dependent linear transformation on the field
\begin{eqnarray}
\nonumber
\hat \Phi(p,\vec q) &=& \int_{\vec q^\prime} \left(e^{M_k(p)}\right)(\vec q,\vec q^\prime) \Phi(p,\vec q^\prime),\\
\hat \Phi^*(p,\vec q) &=& \int_{\vec q^\prime} \Phi^*(p,\vec q^\prime) \left(e^{M_k^\dagger(p)}\right)(\vec q^\prime,\vec q),
\end{eqnarray}
with an exponentiated matrix $e^{M_k(p)}$ with ``indices'' $\vec q,\vec q^\prime$.
This gives the flow equation
\begin{eqnarray}
\nonumber
\partial_k \Gamma_k {\big |}_{\hat\Phi} &=& \partial_k \Gamma_k {\big |}_\Phi - \int_{p,\vec q}\frac{\delta \Gamma_k}{\delta \hat \Phi(p,\vec q)} \partial_k \hat \Phi(p,\vec q){\big |}_\Phi\\
\nonumber
&& -\int_{p,\vec q}(\partial_k \hat \Phi^*(p,\vec q)){\big |}_\Phi\frac{\delta \Gamma_k}{\delta \hat \Phi^*(p,\vec q)}\\
\nonumber
&=& \frac{1}{2} \text{STr} (\Gamma_k^{(2)}+\hat {\cal R}_k)^{-1} (\widehat{\partial_k {\cal R}_k} - \hat {\cal R}_k (\widehat{\partial_k Q^{-1}}) \hat {\cal R}_k)\\
\nonumber
&& -\frac{1}{2} \Gamma_k^{(1)} (\widehat{\partial_k Q^{-1}}) \Gamma_k^{(1)} \\
&& -  \frac{\delta\Gamma_k}{\delta \hat \Phi}(\partial_k M_k) \hat \Phi- \hat \Phi^* (\partial_k M_k^\dagger) \frac{\delta\Gamma_k}{\delta \hat \Phi^*} .
\label{eq:floweqwithvariablechange}
\end{eqnarray}
In the last equation we used an obvious matrix notation and the abbreviations
\begin{equation}
\begin{split}
& \widehat{\partial_k Q^{-1}} = e^{M_k^\dagger} (\partial_k Q^{-1}) e^{M_k}, \quad \hat {\cal R}_k = e^{-M_k^\dagger} {\cal R}_k e^{-M_k}, \\ 
& \widehat{\partial_k {\cal R}_k} = e^{-M_k^\dagger} \partial_k {\cal R}_k e^{-M_k}
\end{split}
\label{eq:abbreviations}
\end{equation}
and functional derivatives are now taken with respect to $\hat \Phi$. It is important to note that the matrices $\partial_k Q$, $\widehat{\partial_k Q}$ and $M_k$ have entries only in the $\Phi$-$\Phi$-block. For example, Eq.\ \eqref{eq:abbreviations} implies a transformation of the cutoff term for the composite bosons but not for the fundamental fields $\psi_1, \psi_2$.
For simplicity we drop the hats at most places below. 

To investigate the implications of the flow equation \eqref{eq:floweqwithvariablechange} we use again a truncation of the flowing action in terms of a vertex expansion. Due to the nonrelativistic dispersion relation and the instantaneous interaction, this leads to exact flow equations for the considered $n$-point functions since their flow equations decouple from higher vertex functions. More concrete, we choose as our truncation
\begin{eqnarray}
\nonumber
\Gamma_k &=& \int_p \left\{\psi_1^*\left(p_0-\frac{1}{2M_1}\vec p^2\right)\psi_1+\psi_2^*\left(p_0-\frac{1}{2M_2}\vec p^2\right)\psi_2\right\}\\
\nonumber
&& - \int_{p,\vec q,\vec q^\prime} (\psi_1\psi_2)^*(p,\vec q) V(\vec q- \vec q^\prime) (\psi_1\psi_2)(p,\vec q^\prime)\\
\nonumber
&& - \int_{p,\vec q,\vec q^\prime} (\psi_1\psi_2)^*(p,\vec q) \lambda_\psi(p,\vec q, \vec q^\prime) (\psi_1\psi_2)(p,\vec q^\prime)\\
\nonumber
&& - \int_{p,\vec q,\vec q^\prime} \left\{ \Phi^*(p,\vec q) h_\Phi(p,\vec q, \vec q^\prime) (\psi_1\psi_2)(p,\vec q^\prime)+c.c. \right\}\\
&& + \int_{p,\vec q, \vec q^\prime} \Phi^*(p,\vec q) P_\Phi(p,\vec q,\vec q^\prime) \Phi(p,\vec q^\prime).
\label{eq:truncationrebosonization}
\end{eqnarray}
Once again, the propagators for the fields $\psi_1, \psi_2$ remain unmodified by the renormalization flow, while the functions $\lambda_\psi$, $h_\Phi$ and $P_\Phi$ are $k$-dependent objects. For $k=\Lambda$ they have the initial values
\begin{eqnarray}
\nonumber
\lambda_{\psi,\Lambda}(p,\vec q,\vec q^\prime) &=& 0,\\
\nonumber
h_{\Phi,\Lambda} (p,\vec q,\vec q^\prime) &=& V(\vec q-\vec q^\prime),\\
P_{\Phi,\Lambda} (p,\vec q,\vec q^\prime) &=& Q_\Lambda(\vec q,\vec q^\prime) = \infty\; \delta^{(3)}(\vec q-\vec q^\prime).
\end{eqnarray}
Projecting the flow equation \eqref{eq:floweqwithvariablechange} onto the truncation in Eq.\ \eqref{eq:truncationrebosonization} we find for $\lambda_\psi=0$ the flow equations (again in matrix notation and suppressing the argument $p$ at several places)
\begin{eqnarray}
\nonumber
\partial_k \lambda_\psi &=& \tilde \partial_k (V A_k^{-1} V) + h_\Phi (\partial_k Q^{-1}) h_\Phi,\\
\nonumber
\partial_k h_\Phi &=& \tilde \partial_k (h_\Phi A_k^{-1} V) - P_\Phi (\partial_k Q^{-1}) h_\Phi\\
\nonumber
&&- (\partial_k M_k^\dagger) h_\Phi,\\
\nonumber
\partial_k P_\Phi &=& -\tilde \partial_k (h_\Phi A_k^{-1} h_\Phi) - P_\Phi (\partial_k Q^{-1}) P_\Phi\\
&& - (\partial_k M_k^\dagger) P_\Phi - P_\Phi (\partial_k M_k).
\label{eq:floweq234}
\end{eqnarray}
As discussed in ref.\ \cite{FW09} we can now use our freedom in the choice of $\partial_k Q^{-1}$ to enforce $\partial_k \lambda_\psi=0$, i.e.
\begin{equation}
\lambda_\psi(p,\vec q, \vec q^\prime)=0 \quad \text{for all } k.
\end{equation}
This fixes $\partial_k Q^{-1}=-h_\Phi^{-1} V (\partial_k A_k^{-1}) V h_\Phi^{-1}$. In addition, the $k$-dependent field rescaling determined by the matrix $M_k(p)$ can be chosen arbitrary as well such that we can use it to enforce $\partial_k h_\Phi(p)=0$, i.e.
\begin{equation}
h_\Phi(p,\vec q, \vec q^\prime)=V(\vec q-\vec q^\prime) \quad \text{for all } k.
\end{equation}
In summary, we use
\begin{eqnarray}
\nonumber
\partial_k Q^{-1} &=& -\partial_k A_k^{-1},\\
\nonumber
\partial_k M_k &=& (\partial_k A_k^{-1})^\dagger(P_\Phi+V)^\dagger,\\
\partial_k M_k^\dagger &=& (P_\Phi+V) (\partial_k A_k^{-1}).
\label{eq:dkQdkM}
\end{eqnarray}
For $P_\Phi$ this leads to the flow equation
\begin{eqnarray}
\nonumber
\partial_k P_\Phi &=& - V (\partial_k A_k^{-1}) V + P_\Phi (\partial_k A_k^{-1}) P_\Phi\\
\nonumber
&& - (P_\Phi + V)(\partial_k A_k^{-1}) P_\Phi - P_\Phi (\partial_k A_k^{-1})(P_\Phi+V)\\
&=& - (P_\Phi+V)(\partial_k A_k^{-1})(P_\Phi+V)
\end{eqnarray}
which is solved by
\begin{equation}
P_\Phi = A_k - V.
\end{equation}
Using the definition of $A_k$ in Eq.\ \eqref{eq:akinverse}, we see that, as expected, the zero crossings of $P_\Phi$ at $k=0$ correspond to the solution of Schr\"odingers equation for the two particle problem. Note that we can now also determine via Eqs. \eqref{eq:dkQdkM} and \eqref{eq:abbreviations} the $k$-dependent matrix $Q$ in the Hubbard-Stratonovich transformation \eqref{eq:Spb2}. 

To summarize, we found the solution to the flow equation by first adjusting the scale-dependent Hubbard-Strat\-on\-ovich transformation $\partial_k Q^{-1}$ such that no additional term $\lambda_\psi$ is generated by the flow. All contributions to an interaction of this form are dynamically expressed by bound state exchange processes. As a second step we choose a $k$-dependent linear field redefinition (encoded by the matrix $M_k$) such that also the Yukawa interaction $h_\Phi$ remains $k$-independent. While the first step (to enforce vanishing $\lambda_\psi$) is rather natural, the second is more arbitrary. In our case the choice of $M_k$ was guided by the insight we obtained in section \ref{sec:Partialbosonization}. More general it is useful to redefine the composite fields such that the Yukawa-like interaction $h_\Phi$ remains independent of the center of mass momentum $p$, which is always possible. For some problems and for suitable choices of the cutoff function $R_k$ it may also be possible to choose the basis for the composite fields $\Phi$ such that the inverse propagator $P_\Phi$ is a diagonal matrix. As discussed in section \ref{sec:Partialbosonization} this is possible for the problem considered here for very simple cutoff functions such as $R_k=k^2$. For more complicated $R_k$ one can still diagonalize $P_\Phi$ but the drawback is then that $h_\Phi$ might depend on $p$.

\section{Approximative solutions}
\label{sec:approximationschemes}

Up to this point we discussed the application of the functional RG to the nonrelativistic two-body problem in a rather formal way. No approximations were needed, but we had to introduce an additional functional integral over a bilocal composite fields $\Phi(x,\vec y)$ where $\vec y$ labels the relative coordinate of the two particles. For more complicated problems it is difficult to follow the RG flow for bilocal fields. It is therefore necessary to find useful approximation schemes. One possibility is to expand the field in terms of an complete orthonormal set of functions $f_n(\vec y)$ like
\begin{equation}
\Phi(x,\vec y) = \sum_{n\in {\cal I}} \phi_n(x) f_n(\vec y).
\end{equation}
One can then consider the flow of the propagator and coupling constants in that basis. A sensible approximation would now be to take only a finite subset ${\cal J}$ of the infinite index set ${\cal I}$ into account and to neglect all couplings between the fields $\phi_n$, $n\in {\cal J}$ and the fields $\phi_n$, $n\notin {\cal J}$. The approximation becomes good if the influence of the neglected couplings onto the flow of the considered quantities is small. This will, of course, only work for some particular choices of the set $f_n(\vec y)$.

In the following we will consider as an example a Yukawa potential of the form
\begin{equation}
V(\vec x-\vec y) = \frac{-e^2}{4\pi |\vec x_1-\vec x_2|} e^{-m|\vec x_1-\vec x_2|}.
\label{eq:Yukawapot}
\end{equation}
Note that for $m\to \infty$ this potential approaches a contact interaction
\begin{equation}
V(\vec x-\vec y) \to -\frac{e^2}{m^2} \delta^{(3)}(\vec x-\vec y).
\label{eq:Yukawatocontact}
\end{equation}
We choose a particular simple set of orthonormal functions
\begin{equation}
\Phi(x,\vec y) = \sum_{n,l,m} \phi_{nlm} (x)\; Y_{lm}(\Omega_{\vec y})\; R_{nl}(|\vec y|).
\label{eq:expansion}
\end{equation}
Here we use the spherical harmonics $Y_{lm}$ and some functions $R_{nl}$ for the radial direction. A possible choice would be the associate Laguerre Polynomials or any other suitable normalized set of orthogonal functions. Depending on the cutoff function, it may be possible to find analytic solutions for the flow equations in a particular basis.

As a test of the robustness of our formalism, we will try a very crude approximation in the following and include only the term with $l=m=0$ in Eq. \eqref{eq:expansion}. Also the radial dependence is truncated to a single function $R(|\vec y|)$ with support only for $|\vec y|=0$. In other words, we use 
\begin{equation}
\Phi(x,\vec y) = \phi(x) \delta^{(3)}(\vec y).
\label{eq:compfieldpointbos}
\end{equation}

The microscopic action corresponding to Eq.\ \eqref{eq:Yukawapot} reads in momentum space
\begin{eqnarray}
\nonumber
S_\psi &=& \int_p{\bigg \{} \psi_1^*(p) \left(p_0-\frac{1}{2M_1}\vec p^2\right)\psi_1(p) \\
\nonumber
&& + \psi_2^*(p) \left(p_0-\frac{1}{2M_2}\vec p^2\right) \psi_2(p) {\bigg \}} \\
\nonumber
&& + \int_{q_1..q_4} \psi_2^*(q_4) \psi_1^*(q_3) \frac{e^2}{m^2+(\vec q_1-\vec q_3)^2} \\
&& \times \psi_1(q_1) \psi_2(q_2) \delta(q_1+q_2-q_3-q_4).
\label{eq:Coulombmomentumspace}
\end{eqnarray}
It is useful to introduce the real auxiliary field $\sigma$ which can be seen as the remnant of a massive photon in the nonrelativistic limit. Similar to the Hubbard-Stratonovich transformation performed in section \ref{sec:Partialbosonization}, we multiply to the partition function a functional integral over the fields $\sigma_1$ and $\sigma_2$ weighted by the quadratic action
\begin{equation}
\begin{split}
S_\sigma = & -\int_p \left(\sigma_1(-p)-\frac{e}{\vec p^2}(\psi_1^* \psi_1)(-p)\right) (m^2+\vec p^2) \\
& \left(\sigma_2(p)-\frac{e}{\vec p^2}(\psi_2^* \psi_2)(p)\right)
\end{split}
\label{eq:sigmaHST}
\end{equation}
with
\begin{equation}
(\psi_1^*\psi_1)(p) = \int_q \psi_1^*(q) \psi_1(q+p)
\end{equation}
and similar for $(\psi_2^*\psi_2)$. Adding this to Eq.\ \eqref{eq:Coulombmomentumspace} we arrive at
\begin{eqnarray}
\nonumber
S_{\psi\sigma} &=& \int_p {\bigg \{}\psi_1^*(p)\left(p_0-\frac{1}{2M_1}\vec p^2\right)\psi_1(p)\\
\nonumber
&& +\psi_2^*(p)\left(p_0-\frac{1}{2M_2}\vec p^2\right)\psi_2(p)\\
\nonumber
&&+ \sigma_1(-p) (-m^2-\vec p^2) \sigma_2(p) {\bigg \}}\\
&&+e\int_{p_1,p_2} {\big \{} \sigma_1(p_1-p_2) \psi_2^*(p_1) \psi_2(p_2)\\
&&+\sigma_2(p_1-p_2) \psi_1^*(p_1) \psi_1(p_2){\big \}}.
\end{eqnarray}
For the flowing action $\Gamma_k$ we make the following truncation
\begin{equation}
\begin{split}
& \Gamma_k[\psi,\phi,\sigma] = S_{\psi\sigma} + \int_p \phi^*(p) P_\phi(p) \phi(p)\\
& - \int_{p,q} \left\{\phi^*(p) h_\phi(p)\psi_1(\eta_1 p+q)\psi_2(\eta_2 p-q)+c.c.\right\}\\
&- \int_{p,q_1,q_2} \psi_2^*(\eta_2 p-q_1) \psi_1^*(\eta_1 p+q_1)\\
& \times \lambda_\psi(p)\, \psi_1(\eta_1 p+ q_2) \psi_2 (\eta_2 p-q_2).
\end{split}
\label{eq:truncationsingleboson}
\end{equation}
We will use our freedom in the choice of the scale-dependent Hubbard-Stratonovich transformation to ensure that $\lambda_\psi(p)=0$ at all scales. It is a consequence of the nonrelativistic dispersion relation that the couplings in the $\sigma$-sector of the theory (the charge $e$ and the propagator of the $\sigma$-boson) do not receive any modifications from the renormalization group flow. This can also be checked explicitly by looking at the flow equations for these quantities. One advantage of introducing the field $\sigma$ instead of working with $V(\vec q-\vec q^\prime)$ as before is that one can also introduce a cutoff function for $\sigma$. In principle, one could do this also for the composite boson $\phi$, but we will not do this for simplicity, here. We use
\begin{eqnarray}
\nonumber
\Delta S_k &=& \int_p {\bigg \{} \psi_1^*(p) \left[-\frac{1}{2M_1} R_k(\vec p^2)\right]\psi_1(p)  \\
\nonumber
&& + \psi_2^*(p) \left[-\frac{1}{2M_1} R_k(\vec p^2)\right] \psi_2(p) {\bigg \}}\\
&& + \sigma_1(-p) [-R_k(\vec p^2)] \sigma_2(p)
\label{eq:cutoff}
\end{eqnarray}
with $R_k(z)=(k^2-z)\theta(k^2-z)$.

Now that we have fixed the truncation and the cutoff function, we can determine the flow equations. By setting the ansatz in Eq.\ \eqref{eq:truncationsingleboson} into the flow equation \eqref{eq:floweqwithvariablechange} we find in the center of mass frame ($\vec p=0$)
\begin{eqnarray}
\nonumber
\partial_k \lambda_\psi(p) &=& \tilde \partial_k \int_{\vec q} \frac{e^4}{(m^2+\vec q^2+R_k(\vec p^2))^2}\\ 
&&\times \frac{1}{p_0-\frac{1}{2\mu}(\vec q^2+R_k(\vec q^2))}\\
\nonumber
&& + h_\phi(p)^2 \partial_k Q^{-1}(p),\\
\nonumber
\partial_k h_\phi(p) &=& -\tilde \partial_k \int_{\vec q} h_\phi(p) \frac{e^2}{m^2+\vec q^2+R_k(\vec p^2)} \\
&& \times \frac{1}{p_0-\frac{1}{2\mu}(\vec q^2+R_k(\vec q^2))}\\
\nonumber
&&-P_\phi(p)  h_\phi(p) \partial_k Q^{-1}(p) - h_\phi(p) \partial_k M(p),\\
\nonumber
\partial_k P_\phi(p) &=& -\tilde \partial_k \int_{\vec q} h_\phi(p) \frac{1}{p_0-\frac{1}{2\mu}(\vec q^2+R_k(\vec q^2))} h_\phi(p)\\
&&- P_\phi(p)^2 \partial_k Q^{-1}(p) - 2 h_\phi(p) \partial_k M(p).
\end{eqnarray}
Here, the derivative $\tilde \partial_k$ hits only the explicit $k$-dependence of the cutoff function $R_k$. Before we solve the flow equations it remains to perform the integration over the spatial momentum $\vec q$. For the cutoff function in Eq.\ \eqref{eq:cutoff} the integration is very simple and gives
\begin{eqnarray}
\nonumber
\partial_k \lambda_\psi(p) &=& \frac{e^4 k^3}{6\pi^2} \partial_k \frac{1}{(m^2+k^2)^2(p_0-\frac{1}{2\mu}k^2)}\\
&& + h_\phi(p)^2 \partial_k Q^{-1}(p),\\
\nonumber
\partial_k h_\phi(p) &=& -\frac{e^2 h_\phi(p) k^3}{6\pi^2} \partial_k \frac{1}{(m^2+k^2)(p_0-\frac{1}{2\mu}k^2)}\\
&&-P_\phi(p)  h_\phi(p) \partial_k Q^{-1}(p) - h_\phi(p) \partial_k M(p),\\
\nonumber
\partial_k P_\phi(p) &=& -\frac{h_\phi(p)^2 k^3}{6\pi^2}\partial_k \frac{1}{p_0-\frac{1}{2\mu}k^2}\\
&&- P_\phi(p)^2 \partial_k Q^{-1}(p) - 2 P_\phi(p) \partial_k M(p).
\end{eqnarray}
We choose now the scale-dependent Hubbard-Stratonovich transformation $\partial_k Q^{-1}(p)$ such that $\partial_k \lambda_\psi(p)=0$. This gives
\begin{equation}
\partial_k Q^{-1}(p) = -\frac{e^4 k^3}{6\pi^2 h_\phi(p)^2} \partial_k \frac{1}{(m^2+k^2)^2(p_0-\frac{1}{2\mu}k^2)}.
\end{equation}
In addition, we choose the rescaling of the field $\phi(p)$ encoded by $\partial_k M(p)$ such that $\partial_k h_\phi(p)=\partial_k h_\phi(0)$. What remains then is the freedom to choose $\partial_k M(0)$. This is just the freedom to make a $p$-independent rescaling of the field $\phi$. Usually, one chooses this wavefunction renormalization such that the propagator for the field $\phi$ has a residue of value unity at a pole that corresponds to a propagating particle. However, since $\partial_k M(0)$ can be chosen arbitrary in principle, we choose it here for simplicity such that $\partial_k h_\phi(0)=0$. This results in
\begin{eqnarray}
\partial_k M(p) &=& -\frac{e^2 k^3}{6\pi^2} \partial_k \frac{1}{(m^2+k^2)(p_0-\frac{1}{2\mu}k^2)} \\
\nonumber
&&+ \frac{e^4 k^3}{h_\phi^2 6\pi^2} P_\phi(p) \partial_k \frac{1}{(m^2+k^2)^2(p_0-\frac{1}{2\mu}k^2)}.
\end{eqnarray}
We keep in mind, that with this choice the poles in the propagator for the field $\phi$ may have residues with values different from one. To summarize, we have obtained the flow equations
\begin{eqnarray}
\partial_k \lambda_\psi(p) &=& 0,\\
\partial_k h_\phi(p) &=& 0,\\
\nonumber
\partial_k P_\phi(p) &=& -\frac{h_\phi^2 k^3}{6\pi^2} \partial_k \frac{1}{p_0-\frac{1}{2\mu} k^2}\\
\nonumber
&+& 2 P_\phi(p) \frac{e^2 k^3}{6\pi^2} \partial_k \frac{1}{(m^2+k^2)(p_0-\frac{1}{2\mu}k^2)}\\
&-& P_\phi(p)^2 \frac{e^4 k^3}{h_\phi^2 6\pi^2} \partial_k \frac{1}{(m^2+k^2)^2(p_0-\frac{1}{2\mu}k^2)}.
\label{eq:flowbder}
\end{eqnarray}

Before we proceed with the solution of the flow equation let us discuss the initial conditions for large values of the scale $k\to\Lambda$. In the limit $k\to\Lambda\to\infty$ the contribution of the $\phi$-particle exchange to the effective interaction between the $\psi$-particles should vanish which implies
\begin{equation}
\lim_{\Lambda\to\infty} \frac{h_{\phi,\Lambda}^2}{P_{\phi,\Lambda}}=0.
\end{equation}
For large but finite $\Lambda$ this value gets modified by one-loop contributions which can be obtained from the flow equation for $\lambda_\psi$
\begin{eqnarray}
\nonumber
\frac{h_{\phi,\Lambda}^2}{P_{\phi,\Lambda}} &=& \int_{\vec q} \frac{e^4}{(m^2+\vec q^2+R_k(\vec q^2))^2} \frac{1}{p_0-\frac{1}{2\mu}(\vec q^2+R_\Lambda(\vec q^2))}\\
&\to& -\frac{2\mu e^4}{3\pi^2 \Lambda^3}.
\end{eqnarray}
In the last line we assumed $\Lambda^2 \gg m^2$, $\Lambda^2\gg 2\mu p_0$. Indeed we find that this vanishes in the limit $\Lambda\to\infty$. Since only the above ratio is fixed, there is still some ambiguity in the choice of the initial conditions. Indeed it is always possible to rescale the field $\phi$ which changes $h_{\phi,\Lambda}^2$ and $P_{\phi,\Lambda}$ but keeps the ratio fixed. Physical observables are of course not affected by such a rescaling. For simplicity we choose
\begin{eqnarray}
\nonumber
h_{\phi,\Lambda} &=& h_\phi = e,\\
P_{\phi,\Lambda} &=& -\frac{3\pi^2}{2\mu e^2} \Lambda^3 = -\Lambda^3/c.
\label{eq:initialcond}
\end{eqnarray}
The last equation defines the abbreviation $c$.
We have now all ingredients to solve the flow equation. From Eq.\ \eqref{eq:flowbder} we obtain
\begin{equation}
\begin{split}
\partial_k P_\phi = & -c \frac{1}{(1-2\mu p_0/k^2)^2}\\
& +\frac{2 c P_\phi}{k^2} {\bigg [}\frac{1}{(1-2\mu p_0/k^2)^2(1+m^2/k^2)}\\
& +\frac{1}{(1-2\mu p_0/k^2)(1+m^2/k^2)^2} {\bigg ]}\\
& -\frac{c P_\phi^2}{k^4} {\bigg [ } \frac{1}{(1-2\mu p_0/k^2)^2(1+m^2/k^2)^2}\\
& +\frac{2}{(1-2\mu p_0/k^2)(1+m^2/k^2)^3}{\bigg ]}.
\end{split}
\label{eq:floweqader}
\end{equation}
We start our investigation with $p_0=0$ and $m^2=0$. The flow equation \eqref{eq:floweqader} simplifies to
\begin{equation}
\partial_k P_\phi = -c + \frac{4c P_\phi}{k^2} - \frac{3c P_\phi}{k^4}.
\end{equation}
The solution of this flow equation with the initial condition in Eq.\ \eqref{eq:initialcond} is quite simple. It is given for $\Lambda\to\infty$ by
\begin{equation}
P_\phi=-k^3/c+k^2.
\end{equation}
We note that the propagator for the composite field $\phi$ changes its sign at the scale $k=c$. The zero crossing of the iinverse propagator corresponds to a bound state. Would we solve the flow equation for a negative frequency $p_0<0$, would the terms involving $p_0$ in Eq.\ \eqref{eq:floweqader} act as an infrared cutoff and effectively stop the flow at the scale where $k^2\approx 2\mu |p_0|$. Choosing $|p_0|\approx c^2/(2\mu)$ leads to $P_\phi=0$ at the macroscopic scale $k=0$ which corresponds to a pole in the propagator for the composite field $\phi$. We note that $c^2/(2\mu)$ is proportional to the Rydberg energy $\sim e^4 \mu$ and the bound state can therefore be interpreted as the lowest bound state of the Coulomb potential. That this comes out qualitatively correct is encouraging, although quantitative agreement with the expectations from quantum mechanics can not be expected in this case due to the simple approximation made in Eq.\ \eqref{eq:compfieldpointbos}.

Let us now come to the limit of very large $m^2$. The potential becomes now of the singular form in Eq. \eqref{eq:Yukawatocontact}. For that reason, the theory must be regularized by introducing a finite UV cutoff $\Lambda_\text{UV}$. However, we will see that many results do not depend on the precise choice of $\Lambda_\text{UV}$. The initial values of couplings are fixed by imposing $h_{\phi,\Lambda}^2/P_{\phi,\Lambda}=0$ at $\Lambda=\Lambda_\text{UV}$. In addition we choose $h_\phi=e$ as before. The flow equation \eqref{eq:floweqader} becomes for large $m^2$
\begin{equation}
\partial_k P_\phi = \left(-c+\frac{2cP_\phi}{m^2}-\frac{cP_\phi^2}{m^4}\right)\frac{1}{(1+w/k^2)^2}
\end{equation}
with the abbreviation $w=-2\mu p_0$. It has the solution
\begin{equation}
\begin{split}
& P_\phi =  m^2  \\
&- 2m^4(k^2+w) {\bigg (}-ck(2k^2+3w)\\
& +(k^2+w)\left[3c\sqrt{w}\;\text{arctan}(k/\sqrt{w})+D\right]{\bigg )}^{-1},
\end{split}
\end{equation}
with an arbitrary constant $D$. The condition that $e^2/P_{\phi,\Lambda}=0$ fixes
\begin{equation}
D=\frac{c\Lambda(2\Lambda^2+3w)}{\Lambda^2+w}-3c\sqrt{w} \; \text{arctan}(\Lambda/\sqrt{w}).
\end{equation}
We note that for large $k$ but $k<\Lambda$ the inverse propagator $P_\phi$ is negative. Depending on the choice of the parameters $m^2$ and $e^2$ (or $c$) $P_\phi$ may change its sign during the flow towards small values of $k$. This corresponds then again to a bound state. To see this we consider $P_\phi$ for $k=0$ which becomes for $\Lambda^2\gg w$
\begin{equation}
P_\phi(k=0)=m^2-\frac{m^4}{c\left(\Lambda-\frac{3\pi}{4}\sqrt{w}\right)}.
\label{eq:sollargemk0}
\end{equation}
To interpret this expression let us assume in the following that the fields $\psi_1$ and $\psi_2$ describe fermions with equal mass $M_1=M_2=M$, $\mu=M/2$. We can obtain the scattering length $a$ from Eq.\ \eqref{eq:sollargemk0} as
\begin{equation}
a=\frac{M}{4\pi}\lambda_{\psi,\text{eff}},
\label{eq:alambda}
\end{equation}
with
\begin{equation}
\lambda_{\psi,\text{eff}} = -\frac{e^2}{m^2}+\frac{e^2}{P_\phi(k=0,w=0)}=\frac{e^2}{c\Lambda-m^2}.
\end{equation}
For a more detailed explanation of Eq.\ \eqref{eq:alambda} we refer to refs.\ \cite{DKS,MFSW}.
Inserting the abbreviation $c=2\mu e^2/(3\pi^2)$ this gives
\begin{equation}
a=\frac{1}{\frac{4\Lambda}{3\pi}-\frac{4\pi m^2}{e^2 M}}.
\end{equation}
Depending on the choice of the initial parameters $e^2$ and $m^2$ for a given value of $\Lambda$, the scattering length can be both negative and positive. For large positive scattering length one expects the presence of a shallow dimer state in the spectrum \cite{BraatenHammer}. Its binding energy can be calculated from the on-shell condition of the $\phi$-particle which implies $P_\phi(k=0)=0$. From this we find
\begin{equation}
c\left(\Lambda-\frac{3\pi}{4}\sqrt{w}\right)=m^2
\end{equation}
or after reinserting $w=-2\mu p_0 = 2\mu E$
\begin{equation}
E=\frac{1}{M}\frac{1}{a^2}.
\end{equation}
This is precisely the well known relation between the scattering length $a$ and the binding energy $E$ of the shallow dimer that exists for $a>0$ \cite{BraatenHammer}. 

It is interesting that our treatment yields exact results in the limit of large $m$. On the other side this could have been expected, since a quite similar formalism using a $k$-independent bosonization yields exact results for the contact potential, as well \cite{DKS,MFSW,Birse}. In any case it is interesting that the relatively simple approximation including only one boson already yields qualitatively correct results for the Coulomb potential and quantitatively precise results for the contact potential. More elaborate approximations will allow quantitative investigations for a large class of nonrelativistic and instantaneous interactions. Since this is not the purpose of the present paper we will discuss in the last chapter some generalizations of the formalism to problems that are in a certain sense more complicated than the nonrelativistic few-body problem.

\section{Generalized formalism}
\label{sec:generalformalismandapproxschemes}

While most parts of this paper where devoted to the treatment of nonrelativistic particles with instantaneous interactions, we will use the present section to present some generalizations of the formalism to a wider class of problems. For this purpose we first somewhat generalize the notion of a composite field. For some set of functions $f_n(y)$ where $y=(y_0,\vec y)$ is a space-time coordinate and for two real numbers $\eta_1,\eta_2\in(0,1)$ with $\eta_1+\eta_2=1$ we define
\begin{eqnarray}
\nonumber
(\psi_1\psi_2)(x,y) &=& \psi_1(x+\eta_2 y) \psi_2(x-\eta_1 y),\\
(\psi_1\psi_2)_n(x) &=& \int_y f_n^*(y) (\psi_1\psi_2)(x,y).
\end{eqnarray}
We assume that the functions $f_n(y)$, $n\in{\cal N}$ constitute a complete orthonormal set such that the following relations hold
\begin{eqnarray}
\nonumber
\int_y f_n^*(y) f_m(y) &=& \delta_{nm},\\
\sum_{n\in{\cal N}} f_n^*(x) f_n(y) &=& \delta(x-y).
\end{eqnarray}
In momentum space with $f_n(y)=\int_p e^{-ipx} \tilde f_n(p)$, $f_n^*(y)=\int_p e^{ipx} \tilde f_n^*(p)$ these relations become
\begin{eqnarray}
\nonumber
\int_p f_n^*(p) f_m(p) &=& \delta_{nm},\\
\sum_{n\in N} f_n^*(p) f_n(q) &=& \delta(p-q),
\end{eqnarray}
and the composite field reads in momentum space
\begin{eqnarray}
\nonumber
(\psi_1\psi_2)_n(p) &=& \int_q \tilde f_n^*(q) \psi_1(\eta_1p+q) \eta_2(\eta_2p-q),\\
(\psi_1\psi_2)_n^*(p) &=& \int_q f_n(q) \psi_2^*(\eta_2p-q) \psi_1^*(\eta_1p + q).
\label{eq:compositefieldsmomspace}
\end{eqnarray}
The index $n$ we use to label the functions $f_n$ is an abstract index that can be either continuous or discrete. Note that the bilocal fields used for the investigation of nonrelativistic particles with instantaneous interactions can be embedded into the above description by choosing $f_n(y)$ to have support only for $y_0=0$, i.e.
\begin{equation}
f_n(y) = \delta(y_0)\; g_n(\vec y).
\end{equation}

Similar to previous chapters we introduce bosonic fields for every composite combination. For example, the term in the action corresponding to the Yukawa-type coupling is
\begin{equation}
\Gamma_k^{(\phi\psi\psi)} = -\sum_{n,m} \int_p \left\{\phi_n^*(p) h_\phi(p)_{nm} (\psi_1\psi_2)_m(p)+c.c.\right\}.
\end{equation}
It will often be possible and convenient to work with a Yukawa coupling that is independent of $p$ and diagonal with respect to the indices $n,m$. The action involves also a term quadratic in the fields $\phi_n^*$, $\phi_n$. We write it as
\begin{equation}
\Gamma_k^{(\phi,2)} = \sum_{n,m} \int_p \phi_m^*(p) P_\phi(p)_{nm} \phi_m(p).
\end{equation}
Similarly, we formally introduce also a quartic interaction term for the fields $\psi$
\begin{equation}
\Gamma_k^{(\psi,4)} = -\sum_{n,m} (\psi_1\psi_2)^*_n(p) \lambda_\psi(p)_{nm} (\psi_1\psi_2)_m(p).
\label{eq:quanricintexpnm}
\end{equation}

As an important side remark, we note that a very large class of quartic interactions can be writte in the form \eqref{eq:quanricintexpnm}. A completely general interaction between the fields $\psi_1$ and $\psi_2$ can be written in a homogeneous situation as
\begin{equation}
\begin{split}
& \Gamma_k^{(\psi,4)} = -\int_{k_1\dots k_4} \psi_2^*(k_2)\psi_1^*(k_1) \;\lambda_\psi(k_1,k_2,k_3,k_4)\\ 
& \psi_1(k_3) \psi_2(k_4)\;\delta(k_1+k_2-k_3-k_4).
\end{split}
\label{eq:lambdapsigeneral}
\end{equation}
Due to momentum conservation, the vertex function $\lambda_\psi$ depends actually only on three independent variables. One possibility to choose them is
\begin{eqnarray}
\nonumber
p &=& k_1+k_2=k_3+k_4,\\
\nonumber
q_1 &=& \eta_2 k_1-\eta_1 k_2,\\
q_2 &=& \eta_2 k_3-\eta_1 k_4.
\end{eqnarray}
Eq.\ \eqref{eq:lambdapsigeneral} can then be written as
\begin{equation}
\Gamma_k^{(\psi,4)} = -\int_{p,q_1,q_2} (\psi_1\psi_2)^*(p,q_1)\;\lambda_\psi(p,q_1,q_2) \; (\psi_1\psi_2)(p,q_2),
\label{eq:explambdapsiqq}
\end{equation}
with
\begin{eqnarray}
\nonumber
(\psi_1\psi_2)(p,q) &=& \psi_1(\eta_1p+q) \psi_2(\eta_2p-q)\\
(\psi_1\psi_2)^*(p,q) &=& \psi_2^*(\eta_2p-q) \psi_1^*(\eta_1p+q).
\end{eqnarray}
Assuming now that $\lambda_\psi(p,q_1,q_2)$ is regular with respect to the arguments $q_1$ and $q_2$ we can expand
\begin{equation}
\lambda_\psi(p,q_1,q_2) = \sum_{n,m} \tilde f_n(q_1) \;\lambda_\psi(p)_{nm} \tilde f_m^*(q_2),
\label{eq:expansionlambdapsi}
\end{equation}
with coefficients
\begin{equation}
\lambda_\psi(p)_{nm} = \int_{q_1,q_2} \tilde  f_n^*(q_1) \; \lambda_\psi(p,q_1,q_2) \; f_m(q_2).
\end{equation}
Plugging Eq.\ \eqref{eq:expansionlambdapsi} into Eq.\ \eqref{eq:explambdapsiqq} and using \eqref{eq:compositefieldsmomspace} we find that this is precisely of the form in Eq.\ \eqref{eq:quanricintexpnm}. Note that we did not assume that $\lambda_\psi(p,q_1,q_2)$ is regular with respect to the argument $p$. Possible singularities are transfered to the coefficients $\lambda_\psi(p)_{nm}$. In fact, it is a big advantage of our formalism that poles or branch cuts in $\lambda_\psi(p)_{nm}$ can be described by relatively simple parameterizations of the inverse propagator $P_\phi(p)_{nm}$ for a composite particle. 

Let us assume that for a fixed Hubbard-Stratonovich transformation a term of the form in Eq.\ \eqref{eq:quanricintexpnm} is generated by the renormalization group flow, i.e. $\partial_k \lambda_\psi(p)_{nm}{\big |}_\text{HS}\neq0$. We can then follow the calculation in Sect.\ \ref{sec:scaledeppartialbos} and use a scale-dependent Hubbard-Stratonovich transformation to realize $\partial_k \lambda_\psi(p)_{nm}=0$. This will lead to additional contributions in the flow equations for $h_\phi(p)_{nm}$ and $P_\phi(p)_{nm}$. In addition, a $k$-dependent linear transformation of the fields $\phi_n(p)$ can be employed, for example to keep $h_\phi(p)_{nm}$ independent of $p$. More concrete, the flow equations can be derived similar to Eq.\ \eqref{eq:floweq234} and read in symbolic notation
\begin{eqnarray}
\nonumber
\partial_k \lambda_\psi(p)_{nm} &=& \partial_k \lambda_\psi(p)_{nm}{\big |}_\text{HS} \\
\nonumber
&&+ \sum_{r,s} h_\phi(p)_{nr} \partial_k Q^{-1}(p)_{rs} h_\phi(p)_{sm},\\
\nonumber
\partial_k h_\phi(p)_{nm} &=& \partial_k h_\phi(p)_{nm}{\big |}_\text{HS} \\
\nonumber
&&- \sum_{r,s} P_\phi(p)_{nr} \partial_k Q^{-1}(p)_{rs} h_\phi(p)_{sm}\\
\nonumber
&&-\sum_r (\partial_k M_k^\dagger)(p)_{nr} h_\phi(p)_{rm},\\
\nonumber
\partial_k P_\phi(p)_{nm} &=& \partial_k P_\phi(p)_{nm} {\big |}_\text{HS} \\
\nonumber
&&- \sum_{r,s} P_\phi(p)_{nr} \partial_k Q^{-1}(p)_{rs} P_\phi(p)_{rm}\\
\nonumber
&&-\sum_r (\partial_k M_k^\dagger)(p)_{nr} P_\phi(p)_{rm}\\
&& - \sum_r P_\phi(p)_{nr} (\partial_k M_k)(p)_{rm}.
\end{eqnarray}

We see that by choosing $\partial_k Q^{-1}(p)_{nm}$ conveniently, we can absorb a quartic interaction term $\lambda_\psi$ as in Eq.\ \eqref{eq:lambdapsigeneral} that is regular with respect to the relative momenta $q_1$ and $q_2$ into the exchange of a composite boson of the type $\psi_1\psi_2$. Besides this particle-particle pair also other combinations are possible, such as the combination $\psi_1^*\psi_1$ or $\psi_2^*\psi_2$ corresponding to particle-hole pairs in nonrelativistic physics or a particle-antiparticle pair in relativistic quantum field theory. With a construction similar to the one presented above, one can show that a contribution to the general interaction term $\lambda_\psi$ in Eq.\ \eqref{eq:lambdapsigeneral} that is regular as a function of $k_1+k_3$ and $k_2+k_4$ (but might have singularities with respect to $k_1-k_3=k_4-k_2$) can be absorbed into the exchange of a $\psi_1^*\psi_1$ pair. The construction would be similar to introducing the $\sigma$-field in Eq.\ \eqref{eq:sigmaHST} but now with a $k$-dependent Hubbard-Stratonovich transformation. In a relativistic field theory (or in a nonrelativistic field theory at nonzero density) the propagator of this field might get a nontrivial frequency dependence and thus become a dynamical field. Finally, the exchange of a $\psi_1^*\psi_2$-pair can describe contributions that are regular as functions of the combinations $k_1+k_4$ and $k_2+k_3$ but might have singularities with respect to $k_1-k_4=k_3-k_2$. 

\section{Conclusions}
\label{sec:conclusions}

We have shown in this paper how a recently derived exact flow equation can be employed for the treatment of bound state formation in quantum field theory. For a nonrelativistic field theory, the relevant flow equations can be integrated exactly and our approach is equivalent to the standard quantum mechanical treatment. The presented formalism can also be employed to find approximate solutions to bound state problems. This can be useful under circumstances where no exact solution from quantum mechanics is available. Examples for such situations arise in relativistic quantum field theory, due to non-instantaneous interactions or in nonrelativistic quantum field theory at nonzero density and temperature.

One big advantage of our formalism is that fundamental and composite particles are described in a very similar way by fluctuating quantum fields. A scale dependent Hubbard-Stratonovich transformation allows to absorb complicated interaction vertices into relatively simple tree diagrams of a Yukawa-type theory. Depending on the situation, a particular composite field might become a propagating degree of freedom, end as a gapped excitation or remain an auxiliary field without own dynamics when the infrared cutoff scale is lowered. 
The formalism is particularly useful in situations where the relevant degrees of freedom change during the renormalization group flow. This is often the case when interaction effects are strong. Prominent examples are QCD where the ultraviolet physics is dominated by quarks and gluons while mesons and baryons dominate the physics in the infrared or the Hubbard model, where fermionic degrees of freedom dominate the UV, while various competing pairing instabilities dominate the infrared physics. 
For the example of QCD the flow from UV to IR degrees of freedom was discussed in a setup closely related to the one discussed here \cite{GiesWetterich,GiesWetterich2,Braun}.

Functional renormalization group equations have been applied with quite some success to many problems in quantum field theory. For reviews see ref. \cite{ReviewRG,Pawlowski,SalmhoferHonerkamp,Metzner}. It is a general feature that attempts to increase the precision of these calculations face the difficult problem to find an efficient parameterization for the momentum dependence of vertex functions. The formalism presented in this paper does not only provide a general framework for such parameterizations (similar to the one proposed in ref.\ \cite{HusemannSalmhofer}) but also allows to take the essential composite degrees of freedom as dynamical fields into account. This includes also possible interactions between composite particles which typically correspond to complicated and nonlocal higher-order interactions in terms of the fundamental fields. Another advantage of the description involving the composite fields is that spontaneous symmetry breaking, for example in a superconducting phase, is straight-forward to describe.

Many of the phenomena discussed above -- including the formation of bound states -- can also be described with other functional methods of QFT such as Schwinger-Dyson and Bethe-Salpeter equations. Renormalization group equations have the advantage that they allow a straightforward discussion of interesting features such as fixed point behavior and universality. Close to renormalization group fixed points one can distinguish between relevant and irrelevant (and marginal) operators. This implies that the flow equations determine which terms in a given expansion become important and which can be neglected! Often, a large class of microscopic theories are attracted towards the same fixed point and the system shows universal features that are independent of the concrete microscopical realization. Using a scale-dependent Hubbard-Stratonovich transformation, one can study for a theory with bound states the different fixed points as well as the crossovers between them \cite{GiesWetterich2}. 

In summary, we have developed a formalism to describe bound states in quantum field theory using an exact flow equation and look forward to applications to many interesting problems in the future.

\acknowledgments
The author thanks C.\ Wetterich for useful discussions. This work has been supported by the DFG research group FOR 723 and the Helmholtz Alliance HA216/EMMI.

\end{document}